\documentstyle[eqsecnum,aps,preprint,tabls,psfig]{revtex}
\newcommand{\st}{\stackrel}
\newcommand{\lh}{\leftrightarrow}
\newcommand{\pp}{\partial}
\newcommand{\be}{\begin{eqnarray}}
\newcommand{\ee}{\end{eqnarray}}
\begin{document}
\tighten
\title{Mesons in Light-Front QCD$_{2+1}$: Investigation of a Bloch 
Effective Hamiltonian}
\author{
{Dipankar Chakrabarti}\thanks{e-mail: dipankar@theory.saha.ernet.in} and
{A. Harindranath}\thanks{e-mail: hari@theory.saha.ernet.in} \\
{\it Saha Institute of Nuclear Physics, 1/AF, Bidhan Nagar, 
	Calcutta 700064 India}}
\date{May 3, 2001}
\maketitle
\begin{abstract}

We study the meson sector of 2+1 dimensional light-front QCD using a Bloch
effective Hamiltonian in the first non-trivial order. The resulting two 
dimensional integral equation is converted into a matrix equation and solved 
numerically. We investigate the efficiency of Gaussian quadrature in achieving
the cancellation of linear and logarithmic light-front infrared divergences. 
The vanishing energy denominator problem which leads to severe 
infrared divergences in 2+1 dimensions is investigated in detail. Our study 
indicates that in the context of Fock space based effective Hamiltonian
methods to tackle gauge theories in 2+1 dimensions, approaches like similarity renormalization
renormalization method may be mandatory  due to uncanceled infrared divergences 
caused by the vanishing energy denominator problem. We define and study 
numerically a reduced model which is relativistic, free from infrared 
divergences, and exhibits logarithmic confinement. The manifestation and  
violation of rotational symmetry as a function of the coupling are studied 
quantitatively. 

\end{abstract}  
\vskip .2in
\vskip .2in

\section{Introduction and Motivation}


There are various well-known motivations\cite{Wilson:1994fk} 
to study QCD in the light-front Hamiltonian formalism. In fact, there have 
been many attempts recently to study relativistic bound state problem in the 
Hamiltonian formalism in a light-front Fock space basis (For a review see, 
Ref. \cite{Brodsky:1998de}). It has been realized that a major impediment to 
a straightforward diagonalization of the Hamiltonian is the rapid growth of 
the dimension of the Hamiltonian matrix with particle number. An alternative 
approach will be to use an {\em effective} Hamiltonian that operates in a few 
particle basis. A challenging problem here is that, for a successful
description of low energy observables, the 
effective Hamiltonian must incorporate main features of strong interaction 
dynamics.

One of the first attempts invoked Tamm-Dancoff\cite{Perry:1990mz} or the
Bloch-Horowitz effective Hamiltonian. Though it was successful in tackling
1+1 dimensional gauge theories, its deficiencies become apparent when
attempts were made in 3+1 dimensions. First and foremost is the lack of
confinement in the case of QCD. 
Second is the appearance of the bound state eigenvalue in the
energy denominators. This has two undesirable consequences. Firstly, a
light-front singularity of the type ${ 1 \over k^+}$, where $k^+$ is the
light front longitudinal momentum of the exchanged gluon, remains in the
bound state equation, which would have canceled if free energies appeared
in the energy denominators. Secondly, from the fermion self energy
contribution, in addition to the mass divergence another ultraviolet
divergence appears (for an example in the context of 3+1 dimensional Yukawa 
model see Ref. \cite{Glazek:1993bs}) 
which contributes to the renormalization of the coupling. This contribution 
is also infrared divergent and can be identified as arising from fermion wave
function renormalization. It is the Fock space truncation that has produced
this unphysical divergence which would otherwise have been canceled by
vertex renormalization in a strict order by order perturbative calculation.  

It is well-known that various standard formulae for effective Hamiltonian all 
have drawbacks. Some of the deficiencies of the Bloch-Horowitz formalism are
absent in the Bloch effective Hamiltonian\cite{Bloch} which was
reinvented in the context of renormalization group by 
Wilson \cite{Wilson:1970tp}.  Bloch Hamiltonian has two desired properties, 
namely, the effective Hamiltonian is (1) Hermitian and (2) involves only 
unperturbed energies in the energy denominator.
Use of Bloch effective Hamiltonian eliminates two major problems of the
Tamm-Dancoff approach to gauge theories mentioned above. However, 
Bloch effective Hamiltonian has an undesirable feature, namely, the
vanishing energy denominator.
To the best 
of our knowledge, Bloch effective Hamiltonian was never assessed in terms of
its strengths and weaknesses  in the study of
bound state problems in field theory.

In the study of bound states, QCD poses challenging problems.
To overcome many pitfalls of standard effective Hamiltonians,
similarity renormalization was proposed\cite{Similar}.  It avoids vanishing
energy denominators and thus provides an improvement over Bloch effective
Hamiltonian. Initial attempts in the similarity renormalization approach
worked in either the non-relativistic limit\cite{Brisudova} or the heavy quark
effective theory context\cite{Zhang:1997dd}. Only recently, work has
begun\cite{Allen:2000kx} to address many practical problems, especially the 
numerical ones one faces in this approach.

%
%
A major feature of gauge theories on the light-front is severe light-front
infrared divergence of the type ${ 1 \over (k^+)^2}$ where $k^+$ is the
exchanged gluon longitudinal momentum which appears in instantaneous
four-fermion, two-fermion two-gluon, and four-gluon interactions. In
old-fashioned perturbation theory these divergences are canceled by
transverse gluon interactions. In similarity perturbation theory the
cancellation is only partial and singular interactions survive. Before
embarking on a detailed study of effective Hamiltonian in the similarity
renormalization approach which is a modification of the Bloch effective 
Hamiltonian, it is quite instructive to study the Bloch effective
Hamiltonian itself. 
The result of such a study can serve as benchmark against which one can
evaluate the merits of similarity renormalization scheme.
This will also provide us quantitative measures on the
strengths and weaknesses of numerical procedures in handling singular
interactions (in the context of light-front field theory) 
on the computer. It is crucial to have such quantitative
measures in order to study the effects of similarity cutoff factors on the
nature of the spectrum. This is one of the motivations for the
present work.     



Just as the Tamm-Dancoff or the Bloch-Horowitz formalism, Bloch effective
Hamiltonian of QCD in the first non-trivial order also does not exhibit
confinement in 3+1 dimensions. Since one of our major concerns is the study
of spectra for confining interactions, we go to 2+1 dimensions. In this case, 
in the limit of heavy
fermion mass, a logarithmic confining potential emerges. 
There are several other reasons also to study light-front QCD in 2+1 
dimensions. They
arise from both theoretical and computational issues which we discuss next.

First of all, issues related to ultraviolet divergence become more
complicated in the light front approach since power counting is
different\cite{Wilson:1994fk}  on the 
light front. We get products of ultraviolet
and infrared divergent  factors which complicate the renormalization
problem. Going to two space one time dimensions greatly simplifies this
issue due to the absence of ultraviolet divergences except in mass
corrections. An extra complication is that Fock space truncation introduces
extra ultraviolet divergences which complicate the situation in
non-perturbative bound state computations\cite{Glazek:1993bs}. Such special
divergences do not occur in 2+1 dimensions. A third complication one
faces in 3+1 dimensions is that on enlarging the Fock space in a bound state
calculation, one soon faces the running of the coupling constant. At low
energy scales, the effective coupling grows 
resulting in a strongly coupled theory\cite{Glazek:1998gt} making the weak
coupling approach with a perturbatively determined Hamiltonian unsuitable or 
making it mandatory to invent mechanisms like non-zero gluon mass to stop the
drastic growth\cite{Wilson:1994fk}. In 2+1 dimensional QCD we do not face this
problem since the coupling constant is dimensionful in this
superrenormalizable field theory and does not run due to ultraviolet
divergence. We {\it can keep} the coupling arbitrarily small and study the
structure of the bound states in a weakly coupled theory. 

Secondly, in 1+1 dimensions, in the gauge 
$A^+=0$, dynamical gluons are absent and
their effect is felt only through instantaneous interactions between
fermions. Further, recall that in light front theory, vacuum is trivial. As
a result, the Fock space structure of the bound states are remarkably
simple. For example, the ground state meson is just a $q {\bar q}$ pair both
at weak and strong couplings. In contrast, in 2+1 dimensions, one component
of the gauge field remains dynamical and one
can systematically study the effects of dynamical gluons. Also note that 2+1
dimensions are the lowest dimensions where glueball states are possible and
offers an opportunity to study their structure in the Fock space language
without additional complications of 3+1 dimensions.


A third reason deals with aspects of rotational symmetry. 
2+1 dimensions offer the first opportunity to investigate violations of
Lorentz invariance introduced by various cutoffs (momenta and/or
particle number) in the context of bound state calculations. This is to be
contrasted with 1+1 dimensions where the sole Lorentz generator, namely
boost, is kinematical in light-front field theory. Since in 2+1 dimensions
we have a superrenormalizable field theory, violations introduced by
transverse momentum cutoffs are minimal. Thus in contrast to 3+1 dimensions, 
one can
study the violations caused by truncation of particle number alone and
longitudinal momentum cutoffs. It is
also conceivable that one can enlarge the Fock space sector and investigate
their effect on restoring Lorentz invariance. It is expected that such
investigations are more viable in 2+1 dimensions compared to 3+1 dimensions
due to less severe demand on computational resources.   



A fourth reason concerns similarity renormalization approach.
In 3+1 dimensions it has been shown that  similarity 
renormalization group
approach\cite{Similar} to effective Hamiltonian in QCD leads to 
logarithmic confining interaction \cite{Perry:1994kp}.  It is of interest 
to investigate corresponding effective
Hamiltonian in  2+1 dimensions especially since the canonical Hamiltonian
already leads to logarithmic confinement in the nonrelativistic limit in
this case. It is also known that in 3+1 dimensions the confining part of the
effective Hamiltonian violates rotational symmetry. Does the violation of
rotational symmetry occur also in 2+1 dimensions? If so, how does it
manifest itself?   

In this work we initiate a systematic study of light-front QCD in
2+1 dimensions to investigate the various issues discussed above.
The plan of the rest of this paper is as follows:
In Sec. II we present the canonical Hamiltonian of 2+1 dimensional QCD. The
Bloch effective Hamiltonian in the $q {\bar q}$ sector in the lowest
non-trivial order 
is derived in Sec. III and the bound state equation is derived. The 
divergence structure is discussed in detail in Sec. IV.
In Sec. V we investigate numerically, cancellation of light-front linear
infrared divergences and the consequences of the vanishing energy
denominator problem which leads to {\em uncanceled} infrared divergences in
the bound state equation. 
A model which is relativistic, free from infrared divergences, and exhibits
logarithmic confinement is presented in Sec. VI. In Sec. VII we present
the numerical investigation of this model in the weak coupling limit.
In this section, we also discuss the violation of rotational symmetry in 
this model at strong coupling. Finally Sec. VIII contains discussion and 
conclusions.  Since Bloch effective
Hamiltonian is unfamiliar to most of the readers, we present a detailed
derivation in Appendix A. Details of the numerical procedures used in this
work are given in Appendix B.
\section{Canonical Hamiltonian}


In this section we present the canonical light front Hamiltonian of 2+1 
dimensional QCD. The Lagrangian density is given by
\be
{\cal L} =\Big [ - { 1 \over 4} (F_{\lambda \sigma a})^2 +
{\overline \psi} ( \gamma^\lambda i D_\lambda - m) \psi \Big ]
\ee
with 
\be
i D^\mu &&= {1 \over 2} \st{\lh}{i\pp^\mu} + g A^\mu, \nonumber \\
F^{\mu \lambda a} && = \partial^\mu A^{\lambda a} - \partial^\lambda A^{\mu
a} + g f^{abc} A^{\mu b} A^{\lambda c}.
\ee
We have the equations of motion,
\be
\Big [ i \gamma^\mu \partial_\mu + g \gamma^\mu A_\mu - m \Big ] \psi &&=0,
\\
\partial_\mu F^{\mu \nu a} + g f^{abc} A_{\mu b} F^{\mu \nu}_c + g {\bar
\psi} \gamma^\nu T^a \psi && =0.
\ee
Because we are in 2+1 dimensions, we immediately face an ambiguity  since
there are no $\gamma$ matrices in 2+1 dimensions. In the literature both
two component \cite{bitar} and four component representation 
\cite{Burkardt:1991xf} have been in use. 
For simplicity, we pick the two component representation. Explicitly,
\be
\gamma^0 =\sigma_2 = \pmatrix{ 0 & -i \cr
                               i & 0 }, ~~\gamma^1 = i \sigma_3 = 
\pmatrix{i & 0 \cr
         0 & -i},~~  \gamma^2 = i \sigma_1 = \pmatrix{0 & i \cr
                                                    i & 0 }.
\ee
\be
\gamma^{\pm} = \gamma^0 \pm \gamma^2,~~ \gamma^+=\pmatrix{ 0 & 0 \cr
                                                          2i & 0}, ~~
\gamma^- = \pmatrix{0 & -2i \cr
                    0 & 0}.
\ee
\be
\Lambda^\pm = { 1 \over 4} \gamma^\mp \gamma^\pm, ~~ \Lambda^+ = \pmatrix{1
& 0 \cr
0 & 0}, ~~ \Lambda^- = \pmatrix{0 & 0 \cr
                                0 & 1}.
\ee
Fermion field operator $ \psi^\pm = \Lambda^\pm \psi$.  We have
\be
\psi^+ = \pmatrix{ \xi \cr
                    0}, ~~ \psi^- = \pmatrix{0 \cr
                                             \eta}
\ee
where $ \xi$ and $\eta$ are one component fields. 
We choose the light front gauge $A^{+a}=0$.
From the equation of motion, we get the equation of constraint
\be
i \partial^+ \psi^- = \Big [ \alpha^1 (i \partial^1 + g A^1)+ \gamma^0 m \Big
] \psi^+.
\ee
Thus the fermion constrained field 
\be
\eta = { 1 \over \partial^+} \Big [ - (i \partial^1 + g A^1) + i m \Big ]
\xi.
\ee
From the equation of motion, in the gauge $A^{+a}=0$, we have the equation 
of constraint
\be
- { 1 \over 2} (\partial^+)^2 A^{-a} = - \partial^1 \partial^+ A^{1a} 
- g f^{abc} A^{1b} \partial^+ A^{1c} - 2 g \xi^\dagger T^a \xi.
\ee
Using the equations of constraint, we eliminate $\psi^-$ and $A^-$ in favor
of dynamical field $\psi^+$ and $A^1$ and arrive at 
the canonical Hamiltonian given by
\be
H = H_0 + H_{int} = \int dx^- dx^1 ({\cal H}_0 + {\cal H}_{int}).
\ee
The free Hamiltonian density is given by
\be
{\cal H}_0 = \xi^\dagger { -(\partial^1)^2 + m^2 \over i \partial^+} \xi
+ { 1 \over 2} \partial^1 A^{1a} \partial^1 A^{1a}.
\ee
The interaction Hamiltonian density is given by
\be
{\cal H}_{int} = {\cal H}_1 + {\cal H}_2
\ee
with
\be
{\cal H}_1 && = g \xi^\dagger A^1{\partial^1 \over \partial^+} \xi + g
\xi^\dagger {\partial^1 \over \partial^+}(A^1 \xi) \nonumber \\
&&~~- g m \xi^\dagger A^1 { 1 \over \partial^+}\xi
+ gm \xi^\dagger {1 \over \partial^+}(A^1 \xi) \nonumber \\
&&~~ -2g {1 \over \partial^+}(\partial^1 A^{1a}) \xi^\dagger T^a \xi 
+ g f^{abc} \partial^1 A^{1a} { 1 \over \partial^+}(A^{1b} \partial^+
A^{1c})
\ee 
and
\be
{\cal H}_2 &&= -2 g^2 \xi^\dagger T^a \xi 
\left({1 \over \partial^+}\right )^2 \xi^\dagger
T^a \xi + g^2 \xi^\dagger A^1 { 1 \over \partial^+} (A^1 \xi) \nonumber \\
&&~~ + 2 g^2 f^{abc} { 1 \over \partial^+} (\xi^\dagger T^a \xi) { 1 \over
\partial^+} (A^{1b} \partial^+ A^{1c}) \nonumber \\
&&~~ + {1 \over 2} g^2 f^{abc} f^{ade} { 1 \over  
\partial^+} (A^{1b} \partial^+ A^{1c}) 
{ 1 \over  
\partial^+} (A^{1d} \partial^+ A^{1e}) . 
\ee
The one component fermion field is given by
\be
\xi(x^+=0, x^-, x^1) = \int {dk^+ dk^1 \over 2 (2 \pi)^2 \sqrt{k^+}}
\Big [  b(k)e^{-ik \cdot x} + d^\dagger(k) e^{ik \cdot x} \Big]. \label{ffe}
\ee
The Fock operators obey the anti commutation relation
\be \{b(k), b^\dagger (q) \} = 2 (2 \pi)^2 k^+ \delta^2(k-q), ~~  
\{d(k), d^\dagger (q) \} = 2 (2 \pi)^2 k^+ \delta^2(k-q),
\ee
other anti commutators being zero.
 Note that in two component representation, 
light front fermions do not carry helicity in 2+1 dimensions.

In free field theory, the equation of motion of the dynamical field
$A^1$ is the same as that of a free 
massless scalar field\cite{Binegar:1982gv} and hence we
can write
\be
A^{1}(x^+=0, x^-, x^1) = \int {dk^+ dk^1 \over 2 (2 \pi)^2 k^+} \Big [
a(k) e^{-ik\cdot x} + a^\dagger(k) e^{ - i k \cdot x} \Big ]. \label{bfe}
\ee
The Fock operators obey the commutation relation
\be
[a(k), a^\dagger(q)] = 2 (2 \pi)^2 k^+ \delta^2(k-q),
\ee
other commutators being zero.

We substitute the Fock expansions, Eqs. (\ref{ffe}) and (\ref{bfe}) into the
Hamiltonian and treat all the terms to be normal ordered. Thus we arrive at
the canonical Hamiltonian in the Fock basis.

\section{Bloch effective Hamiltonian in the meson  
sector and the bound state equation} 

In this section we evaluate the Block effective Hamiltonian to the lowest
non-trivial order for a meson state and derive  the effective bound state
equation. 
We define the $P$ space to be $q {\bar q}$ sector of the Fock space and $Q$
space to be the rest of the space.
In the lowest non-trivial order, the Bloch effective Hamiltonian is given by
(see Appendix A for details)
\be
\langle i \mid H_{eff} \mid j \rangle = \langle i \mid (H_0+H_{int}) \mid j
\rangle +{ 1 \over 2}
\sum_k \langle i \mid v \mid k \rangle \langle k \mid v \mid j \rangle \Big
[{ 1 \over \epsilon_i - \epsilon_k} + { 1 \over \epsilon_j - \epsilon_k}
\Big ].
\ee
The states $ \mid i \rangle$ and $\mid j \rangle$ are, explicitly,
\be
\mid a \rangle &&= b^\dagger(p_1, \alpha) d^\dagger (p_2, \alpha) \mid 0
\rangle , \nonumber \\
\mid b \rangle &&= b^\dagger(p_3, \beta) d^\dagger (p_4, \beta) \mid 0
\rangle,
\ee
where $p_1$, $p_2$ denote momenta and $\alpha$, $ \beta$ denote color which
is summed over. Explicitly, $p_1 = (p_1^+, p_1^1)$ etc.,   where $p_1^+$ is
the plus component and $p_1^1$ is the transverse component. For simplicity of
notation, we will denote the transverse component of momenta without the
superscript 1.  
  
The free part of the Hamiltonian leads to the matrix element
\be \langle a \mid H \mid b \rangle = \left [ {m^2 + p_1^2 \over p_1^+ }
+ {m^2 + p_2^2 \over p_2^+} \right ] 2 (2 \pi)^2 p_1^+ \delta^2(p_1-p_3)
2 (2 \pi)^2 p_2^+ \delta^2(p_2- p_4) \delta_{\alpha \beta} .
\ee
From the four fermion interaction, we get the contribution
\be -4 g^2  (T^a T^a)_{\alpha \alpha} { 1 \over (p_1^+ -
p_3^+)^2} 2 (2 \pi)^2 \sqrt{p_1^+ p_2^+ p_3^+ p_4^+} \delta^2(p_1+p_2- p_3 -
p_4)~ \delta_{\alpha \beta}.
\ee 

Next we evaluate the contribution from the second order term. The
intermediate state $ \mid k \rangle$ is a quark, anti-quark, gluon state. 
This intermediate state gives rise to both self energy and gluon exchange
contributions.  

The self energy contributions are
\be
&& g^2 ~C_f ~\delta_{\alpha \beta} ~p_1^+ 2 (2 \pi)^2 \delta^2(p_1 - p_3)
~p_2^+ 2 (2 \pi)^2 \delta^2(p_2 - p_4) \nonumber \\
&& \int { dk_1^+ dk_1 \over 2 (2 \pi)^2 (p_1^+ - k_1^+)} 
\left \{ -2 {(p_1 - k_1) \over (p_1^+ - k_1^+)} + {k_1 \over k_1^+} + 
{p_1 \over p_1^+} -i { m \over k_1^+} + i {m \over p_1^+} \right \} { 1 \over
ED_1} \nonumber \\
&&~~~~~~~~~~  \left \{ -2 {(p_1 - k_1) \over (p_1^+ - k_1^+)} 
+ {k_1 \over k_1^+} + 
{p_1 \over p_1^+} +i { m \over k_1^+} - i {m \over p_1^+} \right \} \nonumber
\\
&& +g^2 ~C_f ~\delta_{\alpha \beta} ~p_1^+ 2 (2 \pi)^2 \delta^2(p_1 - p_3)
~p_2^+ 2 (2 \pi)^2 \delta^2(p_2 - p_4) \nonumber \\
&& \int { dk_2^+ dk_2 \over 2 (2 \pi)^2 (p_2^+ - k_2^+)} 
\left \{ -2 {(p_2 - k_2) \over (p_2^+ - k_2^+)} + {k_2 \over k_2^+} + 
{p_2 \over p_2^+} -i { m \over k_2^+} + i {m \over p_2^+} \right \} { 1 \over
ED_2} \nonumber \\
&&~~~~~~~~~~  \left \{ -2 {(p_2 - k_2) \over (p_2^+ - k_2^+)} 
+ {k_2 \over k_2^+} + 
{p_2 \over p_2^+} +i { m \over k_2^+} - i {m \over p_2^+} \right \} ,
\ee
with
\be
ED_1 &&= {p_1^2 + m^2 \over p_1^+} - {m^2 + k_1^2 \over k_1^+} - {(p_1 - k_1)^2
\over (p_1^+ - k_1^+)}, \nonumber \\
ED_2 &&= {p_2^2 + m^2 \over p_2^+} - {m^2 + k_2^2 \over k_2^+} - {(p_2 - k_2)^2
\over (p_2^+ - k_2^+)}. 
\ee 
The gluon exchange contributions are
\be
&& -g^2 ~(T^a T^a)_{\alpha \alpha}~ 2 (2 \pi)^2 \delta^2 (p_1 +p_2 - p_3 -
p_4) \sqrt{p_1^+ p_2^+ p_3^+ p_4^+} \nonumber \\
&&~~ \left \{  -2 {(p_1 - p_3) \over (p_1^+ - p_3^+)} + {p_3 \over p_3^+} + 
{p_1 \over p_1^+} -i { m \over p_3^+} + i {m \over p_1^+} \right \} 
\left \{  -2 {(p_1 - p_3) \over (p_1^+ - p_3^+)} + {p_2 \over p_2^+} + 
{p_4 \over p_4^+} +i { m \over p_2^+} - i {m \over p_4^+} \right \}  \nonumber
\\
&&~~{1 \over 2}{\theta(p_1^+ - p_3^+) \over (p_1^+ - p_3^+)} \left \{
{1 \over {m^2 + p_4^2 \over p_4^+} - {(p_1 - p_3)^2 \over (p_1^+ - p_3^+)} -
{m^2 + p_2^2 \over p_2^+}} +  
 {1 \over {m^2 + p_1^2 \over p_1^+} - {(p_1 - p_3)^2 \over (p_1^+ - p_3^+)} -
{m^2 + p_3^2 \over p_3^+}} \right \}\nonumber \\
&& -g^2 ~(T^a T^a)_{\alpha \alpha}~ 2 (2 \pi)^2 \delta^2 (p_1 +p_2 - p_3 -
p_4) \sqrt{p_1^+ p_2^+ p_3^+ p_4^+} \nonumber \\
&&~~ \left \{  -2 {(p_3 - p_1) \over (p_3^+ - p_1^+)} + {p_3 \over p_3^+} + 
{p_1 \over p_1^+} -i { m \over p_3^+} + i {m \over p_1^+} \right \} 
\left \{  -2 {(p_3 - p_1) \over (p_3^+ - p_1^+)} + {p_2 \over p_2^+} + 
{p_4 \over p_4^+} +i { m \over p_2^+} - i {m \over p_4^+} \right \}  \nonumber
\\
&&~~{1 \over 2}{\theta(p_3^+ - p_1^+) \over (p_3^+ - p_1^+)} \left \{
{1 \over {m^2 + p_2^2 \over p_2^+} - {(p_3 - p_1)^2 \over (p_3^+ - p_1^+)} -
{m^2 + p_4^2 \over p_4^+}} +  
 {1 \over {m^2 + p_3^2 \over p_3^+} - {(p_3 - p_1)^2 \over (p_3^+ - p_1^+)} -
{m^2 + p_1^2 \over p_1^+}} \right \}.
\ee


After the construction of $H_{eff}$ in the two particle space, we proceed as
follows. Consider the bound state equation
\be H_{eff} \mid \Psi \rangle = {M^2 +P^2 \over P^+} \mid \Psi \rangle
\ee
where $P^+$, $P$, and $M$ are the longitudinal momentum, the 
transverse momentum
and the invariant mass of the state respectively. The 
state $\mid \Psi \rangle$
is given by 
\be 
\mid \Psi \rangle && = \sum_\beta ~\int {dp_3^+ dp_3 \over \sqrt{2 (2
\pi)^2 p_3^+}}~ \int {dp_4^+ dp_4 \over \sqrt{2 (2
\pi)^2 p_4^+}}~ \phi_2(P; p_3, p_4)~ b^\dagger(p_3, \beta) d^\dagger(p_4,
\beta) \mid 0 \rangle \nonumber \\
&& ~~~~~~~~~~~~~~~~ \sqrt{2 (2 \pi)^2 P^+} \delta^2(P-p_3 -p_4)
\ee
which we symbolically represent as
\be
\mid \Psi \rangle = \sum_j \phi_{2j} \mid j \rangle.
\ee
Taking projection with the state
$ \langle i  \mid  = \langle 0 \mid d(p_2, \alpha) b(p_1, \alpha) $, 
we get the effective bound state equation,
\be {M^2 + P^2 \over P^+} \phi_{2i} = H_{0i} \phi_{2i} + \sum_j \langle i
\mid  H_{Ieff} \mid j \rangle ~ \phi_{2j}.
\ee
Introduce the internal momentum variables $ (x,k)$ and  $(y,q)$ via 
$p_1^+ = xP^+$, $p_1= xP+ k$, 
$ p_2^+ = (1-x)P^+$, $ p_2 = (1-x)P-k$, $p_3^+ = yP^+$, $p_3=yP+q$, $p_4^+ = 
(1-y)P^+$, $ p_4 = (1-y)P -q$ and the amplitude $\phi_2(P; p_1,p_2) ={1
\over \sqrt{P^+}} \psi_2(x, k) $.

The fermion momentum fractions $x$ and $y$ range from 0 to 1. To handle end
point singularities, we introduce the cutoff $ \eta \le x,y \le 1 $. This
does not prevent the gluon longitudinal momentum fraction $x-y$ from
becoming zero and we introduce the regulator $\delta$ such that $ \mid x-y
\mid \ge \delta $. To regulate ultraviolet divergences, we introduce the
cutoff $\Lambda$  on the relative transverse momenta $k$ and $q$. We remind
the reader that in the superrenormalizable field theory under study, only
ultraviolet divergence is in the fermion self energy contribution which we
remove by a counterterm before discretization.

The bound state equation
is
\be
\Big [ M^2 - {m^2 + k^2 \over x (1-x)} \Big ] \psi_2(x,k) &&= SE \
\psi_2(x,k) \ 
- 4 {g^2 \over 2 (2 \pi)^2}C_f \int dy dq ~\psi_2(y,q) ~{ 1 \over (x-y)^2}
\nonumber \\
&& - {g^2 \over 2 (2 \pi)^2}C_f \int dy dq ~\psi_2(y,q) ~{ 1 \over 2}  
{ V \over ED}. \label{ebe1}
\nonumber \\
\ee
The self energy contribution
\be
SE &&=- {g^2 \over 2 (2 \pi)^2 } C_f \int_0^x dy \int  dq
~ xy~ { \Big [ \Big ({q \over y} +{ k \over x} - {2 (k-q) \over (x-y)}
\Big )^2 + {m^2 (x-y)^2 \over x^2 y^2} \Big ] \over
(ky-qx)^2 + m^2 (x-y)^2} \nonumber \\
&&~~- {g^2 \over 2 (2 \pi)^2 } C_f \int_x^1 dy \int  dq
~ (1-x)(1-y) ~ { \Big [ \Big ({q \over 1-y} +{ k \over 1-x} + 
{2 (q-k
) \over (y-x)}
\Big )^2 + {m^2 (y-x)^2 \over (1-x)^2 (1-y)^2} \Big ] \over
[k(1-y)-q(1-x)]^2 + m^2 (x-y)^2} .
\ee 
The boson exchange contribution 
\be
{V \over ED} &&= {\theta (x-y) \over (x-y)} \left [ 
{1 \over {m^2 + q^2 \over y}+ {(k-q)^2 \over (x-y)} - {m^2 + k^2 \over x}} 
+ {1 \over {m^2 + k^2 \over 1-x} + {(k-q)^2 \over x-y} - {m^2 +q^2 \over 1-y}}
\right ] \nonumber \\
&& ~~\times \Big [ K(k,x,q,y) ~ + ~i V_I  \Big ]
\nonumber \\
&& + {\theta (y-x) \over (y-x)} \left [ {1 \over
{m^2 + k^2 \over x}+ {(q-k)^2 \over (y-x)} - {q^2 +m^2 \over y}}+  
{1 \over {m^2 + q^2 \over 1-y} + {(q-k)^2 \over y-x} - {m^2 +k^2 \over 1-x}}
\right ] \nonumber \\
&& ~~\times \Big [ K(q,y,k,x)~ + ~ i V_I  \Big ] ,
\ee
where
\be
K(k,x,q,y) =  \Big ( {q \over y} + {k \over x} - 
2 {(k-q) \over (x-y)} \Big )
\Big ( { q \over 1-y} + { k \over 1-x} + {2 (k-q) \over (x-y)} \Big )
- { m^2 (x-y)^2 \over x y (1-x) (1-y)},
\ee
\be
V_I = - { m \over x y (1-x) (1-y)} [ q (2-y-3x) + k(3y+x-2)].
\ee 
 

\section{Divergence Structure}

In this subsection we carry out a detailed analysis of the divergence
structure of the effective bound state equation. We encounter both infrared
and ultraviolet divergences. 

\subsection{Ultraviolet Divergences}

First consider ultraviolet divergences. In the super renormalizable field
theory under consideration, with the terms appearing in the canonical
Hamiltonian as normal ordered, 
ultraviolet divergence is encountered only in the self energy contributions.
To isolate the ultraviolet divergence, we rewrite 
the self energy integrals as
\be
SE && = - {g^2 \over 2 (2 \pi)^2 } C_f \int_0^x dy \int_{-\Lambda}^{+\Lambda}  dq
   \left [ {(x+y)^2 \over x y (x-y)^2}- { 4 m^2 \over
(ky-qx)^2 + m^2 (x-y)^2} \right ] \nonumber \\
&&~~- {g^2 \over 2 (2 \pi)^2 } C_f \int_x^1 dy \int_{- \Lambda}^{+\Lambda}  dq
\nonumber \\
&&~~~~~~~~~  \left [ {(2 -x -y)^2 \over (y-x)^2 (1-x)(1-y)} -   
{4 m^2  \over
[k(1-y)-q(1-x)]^2 + m^2 (x-y)^2} \right ] .
\ee 
The first term inside the square brackets in the above equation is
ultraviolet divergent, which we cancel by adding an ultraviolet counterterm
given by
\be CT =  + {g^2 \over 2 (2 \pi)^2 } C_f  \int_{-\Lambda}^{+\Lambda} 
 dq \left [\int_0^x dy
    {(x+y)^2 \over x y (x-y)^2} + \int_x^1 dy{(2 -x -y)^2 \over (y-x)^2
(1-x)(1-y)} \right ] . \label{ct} 
\ee
After the addition of this counterterm, the bound state equation
is ultraviolet finite.


\subsection{Infrared Divergences}

The infrared divergences that appear in the bound state equation are of two
types: (1) light front infrared divergences that arise from the gluon
longitudinal momentum fraction $x_g=0$, (2) true infrared divergences that
arise from gluon transverse momentum $k_g=0$ and gluon longitudinal momentum
fraction $x_g=0$. 


\subsubsection{Cancellation of Light-front Infrared Divergences 
in the Effective Bound State Equation}

First consider light front infrared divergences.
The effective bound state equation Eq. ({\ref{ebe1}) explicitly has a linear
light front infrared divergent term ${1 \over (x-y)^2}$ coming from
instantaneous gluon exchange. The most divergent
part of the numerator of the transverse gluon exchange term in this equation is
$-4{(k-q)^2 \over (x-y)^2}$. After combining the terms, the linear infrared
divergent term is completely canceled and the resultant effective bound
state equation takes the form
\be
\Big [ M^2 - {m^2 + k^2 \over x (1-x)} \Big ] \psi_2(x,k) &&= SE1 \
\psi_2(x,k) 
- {g^2 \over 2 (2 \pi)^2}C_f 
\int dy dq ~\psi_2(y,q) \nonumber \\
&&~~~~~~~~~ \times {1 \over 2}\left [ {{\tilde V}_1 \over E_1}
+ {{\tilde V}_2 \over E_2}+ i V_I 
 \left ({1 \over E_1} + { 1 \over E_2} \right ) \right ].  
 \label{ebe2}
\ee 
The self energy contribution, made ultraviolet finite by the addition of the
counterterm is
\be
SE1 && = + {g^2 \over 2 (2 \pi)^2 } C_f \int_0^x dy \int_{-\Lambda}^{+\Lambda}
  dq
   { 4 m^2 \over
(ky-qx)^2 + m^2 (x-y)^2}  \nonumber \\
&&~~ +{g^2 \over 2 (2 \pi)^2 } C_f \int_x^1 dy \int_{- \Lambda}^{+\Lambda}  dq
{4 m^2  \over
[k(1-y)-q(1-x)]^2 + m^2 (x-y)^2} . \label{se1}
\ee
The energy denominator factors are 
\be
{1 \over E_1}   = {x y \over [ky-qx]^2 + m^2 (x-y)^2}, ~~
{ 1 \over E_2} = 
{(1-x)(1-y) \over [k(1-y) - q(1-x)]^2 + m^2 (x-y)^2}.
\ee
The vertex terms are
\be
{\tilde V}_1 = \theta(x-y)  {\tilde U}(k,x,q,y) ~+~ \theta(y-x) {\tilde U}(q,y,k,x) ,
\ee
\be
{\tilde V}_2 && = \theta(x-y) {\tilde U}(k,1-x,q,1-y)  
~+~ \theta(y-x) {\tilde U}(q,1-y,k,1-x) ,
\ee
with
\be
{\tilde U} (k,x,q,y)&& =
4 { m^2 \over x y} - { m^2 (x-y)^2 \over x y (1-x) (1-y)} \nonumber \\ 
&&~~+ {q^2 \over y (1-y)} + {k^2 \over x (1-x)} 
- 2 {k^2 \over (x -y)}{1 \over x (1-x)} + 2 {q^2 \over (x-y)}{1 \over y (1-y)}
\nonumber \\  
&&~~~~+ { kq \over x (1-y)} + {kq \over y (1-x)} + 2 {kq \over (x-y)} \Big [ 
{1 -2 y \over y (1-y)} - {1 - 2 x \over x(1-x)} \Big ].
\ee
In addition to the ${1 \over x_g^2}$ singularity which is canceled, 
transverse gluon exchange 
contributions also contain ${ 1 \over x_g}$ singularity which is removed by 
the principal value prescription. 
Cancellation of this singularity is an appealing feature of the Bloch 
effective Hamiltonian in contrast to the Tamm-Dancoff effective Hamiltonian 
where the singularity cancellation does not occur because of the presence of 
invariant mass in the energy denominator\cite{Krautgartner:1992xz}. 

\subsubsection{$``$True" infrared divergences}


Next we consider true infrared divergences. Consider the self energy
integrals. The energy denominators in these expressions vanish when $k=q$
and $x=y$ which correspond to vanishing gluon momentum. By carrying out the
integrals explicitly, in the limit $ \Lambda \rightarrow \infty $ we get,
\be
SE1 ~= ~{mg^2 \over 2 \pi}~C_f~ \Big [ ~{ 1 \over x}~ 
{\rm ln} \ {x \over \delta}~ +~ {
1 \over 1-x} ~{\rm ln} \ {1-x \over \delta}~ \Big ].
\ee 
Thus the singular part of self energy is 
\be
SE1_{singular} ~= ~- { m g^2 \over 2 \pi}~C_f ~ { 1 \over x (1-x)}~{\rm ln}\
\delta.
\ee 
The infrared divergent contribution from self energy gives a positive
contribution to the fermion mass.
It is important to note that the vanishing of energy denominator is possible
also in 3+1 dimensions, but in that case we do not encounter any divergence.
It is the peculiarity of 2+1 dimensions that the vanishing energy denominators
cause a severe infrared divergence problem.

The same vanishing energy denominators occur also in the one gluon exchange
contributions. Let us now consider various terms in the numerator
separately. The terms proportional to $4m^2$ arose from the denominator of
the transverse gluon exchange. A straightforward calculation shows that this
term leads to both finite and infrared divergent contributions. The infrared
divergent contribution is given by
\be
{ m g^2 \over 2 \pi}~C_f ~ { 1 \over x (1-x)}~{\rm ln}\ \delta
\ee 
which exactly cancels the infrared divergent contribution from self energy.
The finite part, in the nonrelativistic limit, can be shown to give rise to
the logarithmically confining potential. Next we have to consider the
remaining terms in the numerator. Rest of the terms proportional to $m^2$
are multiplied by $(x-y)^2$ so that they do not lead to an infrared
divergence problem. The numerator of the imaginary part vanishes at $k=q$,
and $x=y$ and hence is also infrared finite. It is easy to verify that the
rest of the (transverse momentum dependent) terms in the numerator 
does not vanish when 
the denominator vanishes and hence the resulting bound state equation is
inflicted with infrared divergences arising from the vanishing energy
denominator. This problem was first noted in the context of QED in 2+1
dimensions by Tam, Hamer, and Yung \cite{Tam:1995qk} but was not
investigated by these authors. 
We remind the reader that this is a peculiarity of 2+1
dimensions which provides us a unique opportunity to 
explore the consequences of the vanishing energy denominator problem. 


\section{Numerical study of the bound state equation}

We convert the integral equation into a matrix equation with the use of
Gaussian Quadrature. (For details of the numerical procedure see Appendix
B.) $C_f$ is set to 1 for all the numerical calculations presented.
As mentioned before, an important feature of gauge theories on the
light-front is the presence of linear infrared divergences. They appear in
the canonical Hamiltonian in instantaneous four fermion interaction term.
When the $q {\bar q}g$ states are integrated out {\em completely} in
perturbation theory, they also appear in the effective four fermion
interaction and cancel against each other. Non-cancellation of this
divergence is a major feature of similarity renormalization approach. 
We first address the issue of how linear divergences manifest in the
non-uniform grid of the Gaussian Quadrature and how well it can handle 
linear light front infrared divergence. We have studied numerically discretized
versions of Eq. (\ref{ebe1}) where the divergences are present separately in the
discretized version together with the counterterm given in Eq. (\ref{ct}). 
For $g=.2$, we have calculated the
eigenvalues with and without the instantaneous interaction. The results
presented in Fig. 1(a) for the lowest eigenvalue shows 
that  the Gaussian Quadrature can handle
the cancellation very efficiently.

After the cancellation of linear light-front infrared divergence, a
logarithmic infrared divergence which arises from the vanishing energy
denominator survives in the bound state equation. Here we have to
distinguish two types of terms. First type, where the coefficient of the 
logarithmic infrared divergence is independent of the fermion transverse
momentum and the second type where the coefficient is dependent. Self energy
and Coulomb interaction are of the first type. In the weak coupling limit,
since the wavefunction is dominated by very low transverse momentum, we
anticipate that contributions of the second type will be dynamically
suppressed even though both are multiplied by the same coupling constant.
This is especially true of any discrete grid which automatically imposes a
lower limit on the smallest longitudinal momentum fraction allowed.
Thus at weak coupling, even if there are uncanceled infrared divergences
(divergences of the second type),
they may not be significant numerically whereas divergences of the second
type are significant. By
switching the self energy contribution off and on, we have studied this
interplay. The lowest eigenvalue with and without self energy contribution
is plotted in Fig. 1(b). This shows the cancellation of the 
dominant logarithmic infrared divergence. Since there are still
uncanceled infrared divergences in the bound state equation (with
coefficient proportional to fermion transverse momenta) this figure further
illustrate the fact that such divergences are not numerically significant at
weak coupling.

As the strength of the interaction grows, wavefunction develops medium to
large transverse momentum components and the infrared catastrophe triggered
by the vanishing energy denominator becomes manifest numerically. This is
illustrated in Table I where we present the variation with $\delta$ of the
first five eigenvalues for two different choices of the coupling $g$. The
table clearly shows that on a discrete grid, the uncanceled infrared
divergences due to the vanishing energy denominator problem are not
numerically significant at weak coupling but their effect is readily felt at
a stronger coupling.  

\section{Reduced Model}

In this section we consider a model Hamiltonian free from infrared divergences
constructed by dropping the transverse momentum dependent terms from the
numerator of the effective Hamiltonian. For convenience, we further drop the
terms proportional to $(x-y)^2$ and the imaginary part. This defines our
reduced model which is also ultraviolet finite. The equation governing the
model is given by
\be
\left [ M^2 - {m^2 + k^2 \over x (1-x)} \right ] \psi_2(x,k) &&= SE1 \ 
\psi_2(x,k) \ + \ BE .  
 \label{rbe2}
\ee 
The self energy contribution $SE1$ is the same as given in Eq. (\ref{se1}).
The boson exchange contribution $BE$ is given by
\be
BE && = -{1 \over 2}  {g^2 \over 2 (2 \pi)^2 } C_f \int_0^1 
dy \int_{-\Lambda}^{+\Lambda}
  dq
   { 4 m^2 \over
(ky-qx)^2 + m^2 (x-y)^2} \  \psi_2(y,q) \nonumber \\
&&~~ - {1 \over 2} {g^2 \over 2 (2 \pi)^2 } C_f \int_0^1 
dy \int_{- \Lambda}^{+\Lambda}  dq
{4 m^2  \over
[k(1-y)-q(1-x)]^2 + m^2 (x-y)^2} \ \psi_2(y,q). \label{rme}
\ee


Again we discretize the Eq. (\ref{rbe2}) by Gaussian Quadrature. The convergence
of the eigenvalues as a function of the number of grid points is presented
in Table II. In this table we also present the (in)dependence of eigenvalues on
the momentum cutoff.
2+1 dimensions provide an opportunity to study the manifestation and 
violation of rotational symmetry in light front field theory in a simpler
setting compared to 3+1 dimensions. The absence of spin further facilitates
this study. Rotational symmetry in this case simply implies degeneracy with
respect to the sign of the azimuthal quantum number $l$. Thus we expect all
$ l \neq 0$ states to be two fold degenerate.

By a suitable change of variables, one can easily show that our reduced
model, in the nonrelativistic limit reduces to Schroedinger equation in two
space dimensions with a logarithmic confining potential. 
In the weak coupling limit, since $C_f$ is set to 1, we can compare our results
of the reduced model (where we do not make any nonrelativistic
approximation)
with the spectra obtained in nonrelativistic $QED_{2+1}$. Tam {\it
et al.}\cite{Tam:1995qk} solved the radial Schrodinger equation in momentum 
space for $l=0$ states and Koures \cite{Koures:1995qp} solved the coordinate 
space radial Schrodinger equation
for general $l$. Since we are solving the light front bound state equation,
rotational symmetry is not at all manifest. However, at weak coupling we
expect that the
spectra exhibit rotational symmetry to a very good approximation. 
Our numerical results are compared
with those of Koures in Table III for two values of the coupling. At $g=0.2$ we
find reasonable agreement with the degeneracy in the spectrum. Even at $g=.6$
the violation of rotational symmetry is very small. Splitting of levels
which are supposed to be degenerate become more visible at very strong
coupling as can be seen from Table IV for $g=5$. 

Along with the eigenvalues, the diagonalization process also yields
wavefunctions. We have plotted the wavefunctions corresponding to the first
four eigenvalues in Fig. 2 as a function of $x$ and $k$. All wavefunctions
are normalized to be $ \int_0^1 dx \int dk ~ \psi^2(x,k)=1$. 
The lowest state is
nodeless and corresponds to $l=0$. The next two states correspond to $l=1$
and have one node. It is interesting to note the way the node appears in the
wavefunctions which correspond to degenerate levels. Since the rotational
symmetry cannot be  manifest in the variables $x$ and $k$, how can the
wavefunctions still indicate this? From Fig. 2,  it is clear that 
the way this problem is 
resolved is by
one wavefunction having a node in $k$ and the other wavefunction having a
node in $x$. Thus even if we did not know about the underlying symmetry from
other means, the light-front wavefunctions have a subtle way of indicating
the symmetry.


\section{Summary, Discussion and Conclusions}

In light-front Hamiltonian approach to the bound state problem in gauge
theories, the Bloch effective Hamiltonian has certain advantages compared to
the Tamm-Dancoff or the Bloch-Horowitz formalism. Furthermore, the recently
proposed similarity renormalization approach is a modification of the Bloch
approach. In order to quantitatively estimate the impact of similarity form 
factors in the similarity renormalization approach, it is extremely useful 
to have a quantitative study of the bound state problem in Bloch formalism.  
As far as we
know, Bloch effective Hamiltonian has never been investigated in the context
of the bound state problem in light-front field theory.

To avoid complexities due to ultraviolet divergences we turn to 2+1
dimensions. This allows us to investigate light-front infrared divergences
in the bound state problem in the presence of transverse dynamics without
the additional complication arising from the mixing of ultraviolet and
light-front infrared divergences. Further, 2+1 dimensions allow us to
quantitatively study the manifestation and possible violation of rotational
symmetry in light-front theory in a simpler setting. The emergence of a
logarithmic confining interaction in the limit of heavy fermion masses is an
added impetus to study gauge theories in 2+1 dimensions.    

Only very recently, study of various issues that arise in the
numerical computations in the similarity approach has begun.  
Since similarity renormalization approach is a modification of Bloch
effective Hamiltonian approach, a detailed numerical study of the latter can
serve as benchmark against which one can evaluate the merits of the similarity
approach. It is also important to quantitatively evaluate the strengths and
weaknesses of numerical procedures in handling singular interactions in the
context of light-front dynamics on the computer. 

In this work we have focused on the Gaussian Quadrature (GQ) which is one
straightforward procedure to solve the integral equation by converting it 
into a matrix equation. We have demonstrated the efficiency of the GQ method
in handling linear and logarithmic light-front infrared divergences. 

A major advantage of the similarity approach is that it avoids the vanishing
energy denominator problem that is present in the Bloch formalism. In 2+1
dimensions the vanishing energy denominator leads to severe {\em infrared
divergences} and hence we are presented with a unique opportunity to study
its consequences. We encounter two types of infrared divergences: (1) with a
coefficient proportional to fermion mass and (2) with a coefficient
proportional to fermion transverse momentum. The former type gets canceled
in the bound state equation between fermion dressing by gluon and gluon
exchange between fermions. The latter type is uncanceled which however can
be dynamically suppressed at very weak coupling on a {\em finite} grid. We
have demonstrated that on a discrete grid provided by GQ, the uncanceled
divergences are numerically insignificant at weak coupling whereas the
catastrophe due to their presence is readily felt at stronger coupling.

We proceed to study a reduced model that is free from infrared divergences
and which reduces to the Schroedinger equation with a logarithmic potential
in the nonrelativistic limit. This model provides us an opportunity to study
the simplest manifestation and possible violation of rotational symmetry in
the context of light-front field theory. Even though the Hamiltonian does not
exhibit rotational symmetry we have shown that at weak coupling spectra
exhibit rotational symmetry to a very good approximation. We have also shown
that even though the rotational symmetry is not manifest in the variables $x$
and $k$, light-front wavefunctions have a subtler way of indicating the
underlying symmetry. 

Our study indicates that in the context of Fock space based effective
Hamiltonian
methods to tackle gauge theories in 2+1 dimensions, approaches like similarity renormalization
method are mandatory  due to uncanceled infrared divergences caused by
the vanishing energy denominator problem. It is important to recall that
Bloch effective Hamiltonian is generated by completely integrating out the
intermediate gluons irrespective of whether they are low energy or high
energy. Is this justified in a confining theory?

Now that we have obtained quantitative measures of the vanishing energy
denominator problem and the nature of the spectra at weak coupling of the
Bloch effective Hamiltonian, the next step is to study QCD$_{2+1}$ in the
similarity renormalization approach which avoids the vanishing energy
denominator problem. An important issue here is the nature of new effective
interactions generated by the similarity approach. It has been shown that in
3+1 dimensions, similarity approach generates logarithmic confining
interactions which however breaks rotational symmetry. It is interesting to
investigate the corresponding situation in 2+1 dimensions.          


\appendix


\section{Bloch Perturbation Theory for Effective Hamiltonian}

Bloch perturbation theory was introduced in Ref. \cite{Bloch}. 
Here we follow the treatment in Ref. \cite{Perry:1994mn} 
where the reader can find
many examples of perturbative calculations.

Consider a Hamiltonian $H$ defined at a cutoff $ \Lambda$.
Let us try to lower the cutoff to $ \lambda$. In general, the 
cutoff could be in energy and/or particle number.
Let us denote by
$Q$ the operator that projects on to all of the states removed when the
cutoff is lowered. Let $ P = I - Q$.  We have
\be
Q^2 =Q,~~ P^2=P, ~~ PQ=QP=0.
\ee
Our purpose is to find an effective Hamiltonian $ H_{eff}$ that produces the
same eigenvalues in the sub space $P$ as the original Hamiltonian $H$.

Introduce an
operator $R$ that satisfies
\be
Q \mid \psi \rangle = RP \mid \psi \rangle
\ee
for all eigenstates of the Hamiltonian that have support in the subspace
$P$. $R$ gives the part of $\mid \psi \rangle$ outside the space projected
by
$P$ in terms of the part of $ \mid \psi \rangle$  inside the space. Require
that $R$ gives zero acting on states outside the subspace. This means
$R=RP$, $R=QR$, $R^2=0$. From $R=QR$, we have, $PR=0$.  Note also that
$R^\dagger \neq R$.  

Start from the set of equations
\be
PHP \mid \psi \rangle + PHQ \mid \psi \rangle &&= EP \mid \psi \rangle,
\label{bpp} \\
QHP \mid \psi \rangle + QHQ \mid \psi \rangle &&= EQ \mid \psi \rangle.
\label{bqp}
\ee
From Eq. (\ref{bpp}), 
\be
RPHP \mid \psi \rangle + RPHQRP \mid \psi \rangle = ERP \mid \psi \rangle.
\ee
From Eq. (\ref{bqp}),
\be
QHP \mid \psi \rangle + QHQRP \mid \psi \rangle = ERP \mid \psi \rangle.
\ee
Subtracting,
\be
RH_{PP} - H_{QQ}R + R H_{PQ}R - H_{QP} =0.
\ee
We have introduced the notations, $PHP = H_{PP}$ and so on. 
Put $ H = h + v$ with $[h,Q]=0$. Then
\be
Rh_{PP} - h_{QQ}R - v_{QP} + R v_{PP} - v_{QQ}R + R v_{PQ}R=0
\ee
which shows that $R$ starts first order in $v$.

We start from the eigenvalue equation,
\be
H(P+Q) \mid \psi \rangle = E (P+Q) \mid \psi \rangle.
\ee
i.e., 
\be
H(P+R) P \mid \psi \rangle = E (P+R) P \mid \psi \rangle.
\ee
Multiplying from the left by $(P+R^\dagger)$ we have,
\be
(P+R^\dagger) H(P+R) P \mid \psi \rangle = E (P+R^\dagger)(P+R)P \mid \psi
\rangle.
\ee
Using $ PR=0$, $ R^\dagger P=0$, $
(P+R^\dagger) (P+R) = P + R^\dagger R.$
Thus we can rewrite the eigenvalue equation as
\be
\Big [ { 1 \over 1+ R^\dagger R} \Big ]^{1 \over 2} (P+R^\dagger) H(P+R) 
\Big [ { 1 \over 1+ R^\dagger R} \Big ]^{1 \over 2} [1 + R^\dagger R]^{1
\over
2} P \mid \psi \rangle = E [ 1 + R^\dagger R]^{1 \over 2}P \mid \psi
\rangle.
\ee
i.e.,
\be
H_{eff} \mid \phi \rangle = E \mid \phi \rangle
\ee
where
\be
\mid \phi \rangle = [1+R^\dagger R]^{1 \over 2} P \mid \psi \rangle
\ee
and
\be
H_{eff} = \Big [ { 1 \over 1+ R^\dagger R} \Big ]^{1 \over 2}
(P+R^\dagger)H(P+R) \Big [ { 1 \over 1+ R^\dagger R} \Big ]^{1 \over 2}.
\ee

Our next task is to generate a perturbative expansion. Denote free eigenstates in $P$ by $ \mid a \rangle$, $ \mid b \rangle$, etc.  Denote free
eigenstates in $Q$ by $ \mid i \rangle$, $ \mid j \rangle$, etc.
Then
\be
h_{PP} \mid a \rangle &&= \epsilon_a \mid a \rangle, \nonumber \\
h_{QQ} \mid i \rangle && = \epsilon_i \mid i \rangle.
\ee
Let us compute $R$ to lowest orders in the perturbation theory.
Let us write $ R=R_1+R_2 + \ldots$ where the subscript denotes orders in
$v$. A straightforward calculation leads to
\be
\langle i \mid R_1 \mid a \rangle &&= { \langle i \mid v_{QP} \mid a \rangle
\over \epsilon_a - \epsilon_i}, \\
\langle i \mid R_2 \mid a \rangle &&= - \sum_b { \langle b \mid v \mid a
\rangle  \langle i \mid v \mid b \rangle \over (\epsilon_a - \epsilon_i)
(\epsilon_b - \epsilon_i)} +
\sum_j { \langle i \mid v \mid j
\rangle  \langle j \mid v \mid a \rangle \over (\epsilon_a - \epsilon_i)
(\epsilon_a - \epsilon_j)} .
\ee 
Our next task is to develop a perturbation theory expansion for the
effective Hamiltonian to a given order.

We start from the expression for the effective Hamiltonian. Remember that
$R_1 \sim O(v)$, $R_2 \sim O(v^2)$.

To order $v$, $H_{eff} = PHP$ and hence
\be
\langle a \mid H_{eff} \mid b \rangle = \langle a \mid (h+v) \mid b \rangle.
\ee
To second order in $v$, we have
\be
H_{eff} = [ 1 - { 1 \over 2} R^\dagger R] [ PHP + PHR + R^\dagger HP
+ R^\dagger H R] [ 1 - { 1 \over 2} R^\dagger R].
\ee
From $R^\dagger H R$ we get,
\be
\langle a \mid R^\dagger H R \mid b \rangle = \sum_i \epsilon_i
{ \langle a \mid v \mid i \rangle \langle i \mid v \mid b \rangle \over
(\epsilon_a - \epsilon_i)(\epsilon_b - \epsilon_i)}.
\ee
From $PHR$ and $R^\dagger HP$ terms we get
\be
\sum_i \langle a \mid H \mid i \rangle \langle i \mid R_1 \mid b \rangle+
\sum_i \langle a \mid R_1^\dagger \mid i \rangle \langle i \mid H \mid b
\rangle \\
= \sum_i \Big [ { \langle a \mid v \mid i \rangle \langle i \mid v \mid b
\rangle \over \epsilon_a - \epsilon_i} +
{ \langle a \mid v \mid i \rangle \langle i \mid v \mid b
\rangle \over \epsilon_b - \epsilon_i}.
\ee
From the {\it normalization factors} we get
\be
 - { 1 \over 2} R^\dagger R PHP - { 1 \over 2} PHP R^\dagger R =
- { 1 \over 2} (\epsilon_a + \epsilon_b) \sum_i 
{ \langle a \mid v \mid i \rangle \langle i \mid v \mid b
\rangle \over (\epsilon_a - \epsilon_i)(\epsilon_b - \epsilon_i)} 
\ee
Adding everything, to second order, we have,
\be
\langle a \mid H_{eff} \mid b \rangle = { 1 \over 2} 
\sum_i \langle a \mid v \mid i \rangle \langle i \mid v \mid b \rangle \Big
[{ 1 \over \epsilon_a - \epsilon_i} + { 1 \over \epsilon_b - \epsilon_i}
\Big ].
\ee
If $a=b$, this expression reduces to the familiar second order energy shift.

Why Bloch formalism is preferred over Bloch-Horowitz formalism?

In the former, eigenstates of the effective Hamiltonian are ortho normalized
projections of the original eigenstates. In the latter, they are not.

Consider two ortho normalized eigenstates of the original Hamiltonian $ \mid
\psi_1 \rangle$ and $ \mid \psi_2 \rangle$ with $ \langle \psi_1 \mid \psi_2
\rangle =0$. However, $P \mid \psi_1 \rangle$ and $ P \mid \psi_2 \rangle$
need not be orthogonal, i.e., $ \langle \psi_1 \mid PP \mid \psi_2 \rangle = 
\langle \psi_1 \mid P \mid \psi_2 \rangle \neq 0$.
Consider
\be
\langle \psi_1 \mid \psi_2 \rangle &&= \langle \psi_1 \mid P \mid \psi_2
\rangle + \langle \psi_1 \mid Q^2 \mid \psi_2 \rangle\nonumber \\
&& = \langle \psi_1 \mid P \mid \psi_2 \rangle + \langle \psi_1 \mid
P^\dagger R^\dagger RP \mid \psi_1 \rangle. \label{norm}
\ee
Construct $ \mid {\tilde \psi_1} \rangle = [1+ R^\dagger R]^{1 \over 2} P
\mid \psi_1 \rangle$, $ \mid {\tilde \psi_2} \rangle = [1+ R^\dagger R]^{1
\over 2}P
\mid \psi_2 \rangle$.
 Then
\be
\langle {\tilde \psi_1} \mid {\tilde \psi_2} \rangle = \langle \psi_1 \mid P
\mid \psi_2 \rangle + \langle \psi_1 \mid PR^\dagger R P \mid \psi_2 \rangle
= \langle \psi_1 \mid \psi_2 \rangle.
\ee  

\section{Details of numerical procedure}
{\it Parametrization}: The light-front variables are parametrized in the
following ways in our numerical calculations. The full $k$-interval is
divided into $n1$ quadrature points. 
 $k$ is defined by two
different ways. One definition is
\be 
k={u \Lambda m\over (1-u^2)\Lambda +m},
\label{k1}
\ee
 where $\Lambda$ is the 
ultraviolet cutoff and $u$'s are the quadrature points lying between $-1$
and $+1$, so that $k$ goes from $-\Lambda$ to $+\Lambda$. The other
definition is
\be 
k={1\over \kappa}tan({u \pi\over 2}),
\label{k2}
\ee
 here $\kappa$ is a parameter that can be tuned
to adjust the ultraviolet cutoff. 
The second definition (\ref{k2}) of $k$
is very suitable 
for weak coupling calculations where we need maximum points to be
concentrated near  $k=0$ and get better convergence  than 
 the first definition (\ref{k1}).

 The longitudinal momentum fraction $x$ ranges 
from $0$ to $1$. We divide all $x$- integrations in our calculations  
into two parts, $x$ ranging from 0 to 0.5 and $x$ ranging from 0.5 to 1
and  discretize each $x$-interval into $n2$ quadrature points
 with the
parametrization
\be
x={1+v+2\eta(1-v) \over 4},~~~~~ \eta \le x\le0.5,\\  
x={3+v-2\eta (1+v) \over 4},~~~~~0.5\le x \le 1-\eta,
\ee
where $v$'s are the Gauss-quadrature points lying between $-1$ and $+1$ 
and $\eta(\rightarrow 0)$ is introduced to handle
end-point singularities in $x$ as mentioned before.

To handle the infrared diverging terms we put the
 cutoff $|x-y|\ge \delta$
 and at 
the end we take the limit $\delta\rightarrow 0$. Numerically, it means
that the result should converge as one decreases $\delta$ if there is no net
infrared divergence in the theory.

{\it Diagonalization}: After discretisation, solving the integral equation 
becomes 
a matrix diagonalization problem. The diagonalization has been performed 
by using the packed storage {\it LAPACK}\cite{laug} routines 
{\it DSPEVX} for the reduced model (real symmetric matrix) and {\it ZHPEVX}
for the full Hamiltonian (Hermitian matrix).


\eject
\vskip 1in
\vskip .5in
\figure{\noindent FIG. 1. Cancellation of infrared divergence. Full line
denotes the full Hamiltonian. (a) shows the cancellation 
of light-front infrared divergence by switching on and off the
instantaneous interaction. Filled circles - without instantaneous
interaction. (b) shows the cancellation of logarithmic
infrared divergence by switching on and off the self energy term. Filled
circles - without self energy. The parameters are 
$g=.2$, $\eta=.00001$,
$m=1$, $\kappa=20$, $n_1=40$, $n_2=50$. }
\vskip .3in
\figure {\noindent FIG. 2. The wavefunctions corresponding to the lowest four
eigenvalues of the reduced model as a function of $x$ and $k$. 
The parameters are 
$g=.2$, $\eta=.00001$,
$m=1$, $\kappa=10$, $n_1=46$, $n_2=74$.  (a) Lowest state, (b) first excited state, (c) second
excited state, (d) third excited state. The first and second excited states
should be degenerate in the absence of violation of rotational symmetry.
}  
\eject
\tablinesep=.1in
\arraylinesep=.1in
\extrarulesep=.1in
\begin{tabular}{||c|c|c|c|c|c|c||}
\hline \hline
  $g$  & $\delta$   &  \multicolumn{5}{c||}  {eigenvalues ($M^2$)} \\
\hline
     & 0.00001 & 4.0913 & 4.1113 & 4.1122 & 
4.1181 & 4.1209 \\
\cline{2-7}
     & 0.0001   & 4.0913 & 4.1113 & 4.1122 & 4.1181 &
 4.1209 \\
\cline{2-7}
 0.2  & 0.001   & 4.0913 & 4.1113 & 4.1122 & 4.1181 & 4.1209
\\
\cline{2-7}
     & 0.005    & 4.0901 & 4.1066 & 4.1099 & 4.1100 & 4.1112
\\
\cline{2-7}     
     & 0.01     & 4.0870 & 4.0972 & 4.0972 & 4.0973 & 4.0973
\\
\hline
     & 0.0001   & -187230.4 & -187225.4 & -186664.9 &
 -186664.8 & -31506.9 \\
\cline{2-7}
 0.6 & 0.001   & -187230.4 & -187225.4 & -186664.9 & -186664.8
& -31506.9 \\
\cline{2-7}
     & 0.005   & 1.9094 & 1.9415 & 3.1393 & 3.1399 & 4.5697
\\
\cline{2-7}
     & 0.01    & 4.5735 & 4.7337 & 4.7667 & 4.7832 & 4.8277
\\
\hline
\hline
\end{tabular}
\vskip 1cm
{TABLE I: Variation with $\delta$ of the  full hamiltonian. The parameters
are 
n1=40, n2=50, $\eta$=0.00001, $ \kappa$=20.0 in $k={1 \over \kappa}
tan({u \pi\over 2})$
}
\label{table1}
\eject
\tablinesep=.1in
\arraylinesep=.1in
\extrarulesep=.1in
\vskip 1cm
\begin{tabular}{||c|c|c|c|c|c|c||}
\hline \hline
 n1 &  n2 & 
\multicolumn{5}{c||}   {eigenvalues (lowest five) ($ \kappa $ =10.0)} \\
\hline 
 ~20~ &  ~20~ & ~4.08926~ & ~4.10605~ &  ~4.10768~ & ~4.11061~ &
 ~4.11085~ \\
  ~30~ &  ~30~ & 4.09045 & 4.10909 & 4.11038 & 4.11516 &
 4.11699 \\
  40 &  30 & 4.09045 & 4.10913 & 4.11035 & 4.11524 &
 4.11697 \\
  40  &  40  & 4.09102 & 4.11052 & 4.11154 & 4.11711 &
 4.11951 \\
  40  &  50  & 4.09136 & 4.11133 & 4.11222 & 4.11811 &
 4.12096 \\
  50 &  50 & 4.09136 & 4.11135 & 4.11219 & 4.11816 & 
4.12095 \\
  50  & 60 & 4.09158 & 4.11188 & 4.11263 & 4.12189 &
 4.12290 \\
  46  &  60 & 4.09158 & 4.11187 & 4.11264 & 4.11877 &
 4.12189 \\
  46 &  66 & 4.09168 & 4.11212 & 4.11284 & 4.11905 &
 4.12231 \\
  46 &  74 & 4.09179 & 4.11237 & 4.11305 & 4.11934 &
 4.12276 \\
\hline
n1 &  n2 &   \multicolumn{5}{c||}{eigenvalues (lowest five)    
      ($ \Lambda $=20.0)} \\
\hline
  46 &  74 & 4.09179 & 4.11240 & 4.11301 & 4.11940 &
 4.12273 \\
\hline
\hline
\end{tabular}
\vskip 1cm
{TABLE II: 
 Convergence of eigenvalue with n1 and n2 (reduced model). The parameters
are 
 $m$=1.0, $g$=0.2, $\eta=0.00001$.
   }
\label{table2}
\eject
\tablinesep=.1in
\arraylinesep=.1in
\extrarulesep=.1in
\vskip 1cm
\begin{tabular}{||c|c|c|c|c||}
\hline \hline
 $g$  &              \multicolumn{4}{c||}    {eigenvalues} \\
\hline
    & This   & 4.0918 & (4.1124, 4.1130) & 4.1194 \\ 
0.2    & work   & (4.1227, 4.1235) & (4.1268, 4.1273) &
  (4.1298, 4.1303) \\
\cline{2-5}
    &Koures   & 4.0925~~$(l=0$)& 4.1144~~$(l=1)$&  4.1214~~$(l=0)$ \\
    &(Ref. \cite{Koures:1995qp}) &  4.1260~~($l=2$) &  4.1303~~$(l=1)$ & 4.1340~~$(l=3)$ \\
\hline
    & This   & 4.5856 & (4.7741, 4.7821) & 4.8390 \\
0.6    & work  & (4.8767, 4.8816)& (4.9094, 4.9184) &
 (4.9458, 4.9481) \\
\cline{2-5}
    &Koures  & 4.5806 ~~ $(l=0)$ & 4.7777~~$(l=1)$ & 4.8409~~$(l=0)$ \\ 
    &(Ref. \cite{Koures:1995qp})   & 4.8827~~ $(l=2)$ & 4.9205~~$(l=1)$ &
 4.9545~~ $(l=3)$ \\
\hline
\hline
\end{tabular}
\vskip 1cm
TABLE III: Reduced model. 
The parameters are n1=46, n2=74,  $\eta=0.00001,$  $m$=1.0. 
$ k=tan(q \pi/2)/\kappa,~~ \kappa=20.0$. Eigenvalues within () are 
$\pm l$ degenerate (broken) states.
\label{table3}
\eject
\tablinesep=.1in
\arraylinesep=.1in
\extrarulesep=.1in
\vskip 1cm
\begin{tabular}{||c|c|c|c|c|c|c|c|c||}
\hline \hline
      &                    n1&  n2 & \multicolumn{6}{c||}
      {eigenvalues} \\
\hline
$I$ &   40&  50
&    18.217 & (30.702, 33.499) & 35.206 &
                         (39.955, 41.159) &
 (41.332, 43.271) & (44.134, 45.272) \\
\cline{2-9}                     
            &               46 & 70 &   18.276 & (30.774,
33.616) &
 35.318 & (40.106, 41.331) & (41.483, 43.477) &
                                    (44.375, 45.503) \\
\hline
$II $      &   40 & 50 &    18.980 &
 (31.507, 34.219) & 35.826 & (40.406, 41.888) & (41.921, 43.788) &
(44.345, 45.163) \\
\cline{2-9}
 &                  46 &  70 & 
  19.008 & (31.542, 34.319) & 35.935 & (40.626, 42.031) &
 (42.088, 44.010) & (44.647, 45.780) \\
\hline
\hline
\end{tabular}
\vskip 1cm
{TABLE IV:  First few eigenvalues in the reduced model. The parameters are 
$g$=5.0, $m$=1.0, $\eta$=0.00001. $(I)$ for the parametrization 
$k=u\Lambda m/((1-u^2)\Lambda+m)$, $\Lambda=40.0.$ $(II)$ for the
parametrization $k=tan(u\pi/2)/\kappa$, $\kappa=10.0.$  
\\ Eigenvalues within ( ) are $\pm l$ 
degenerate (broken) states.
    }
\label{table4}
\eject
\begin{figure}
\begin{center}
\psfig{figure=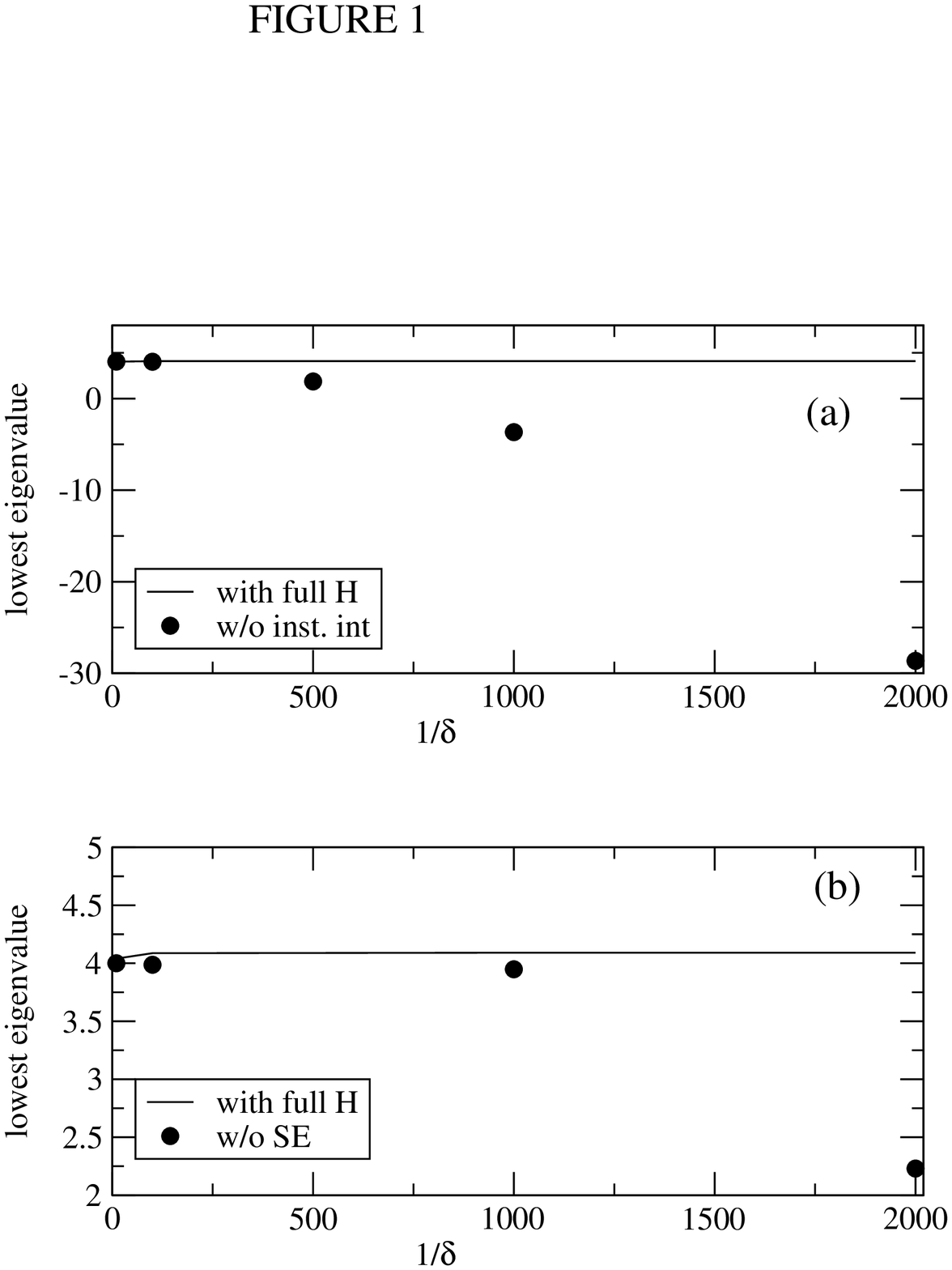,width=16.0cm,height=16.0cm}
\end{center}
\caption{Cancellation of infrared divergence. Full line
denotes the full Hamiltonian. (a) shows the cancellation 
of light-front infrared divergence by switching on and off the
instantaneous interaction. Filled circles - without instantaneous
interaction. (b) shows the cancellation of logarithmic
infrared divergence by switching on and off the self energy term.Filled
circles - without self energy. The parameters are $g=.2$, $\eta=.00001$,
$m=1$, $\kappa=20$, $n_1=40$, $n_2=50$.  
\label{fig1}}. 
\end{figure}
\begin{figure}
\begin{center}
\psfig{figure=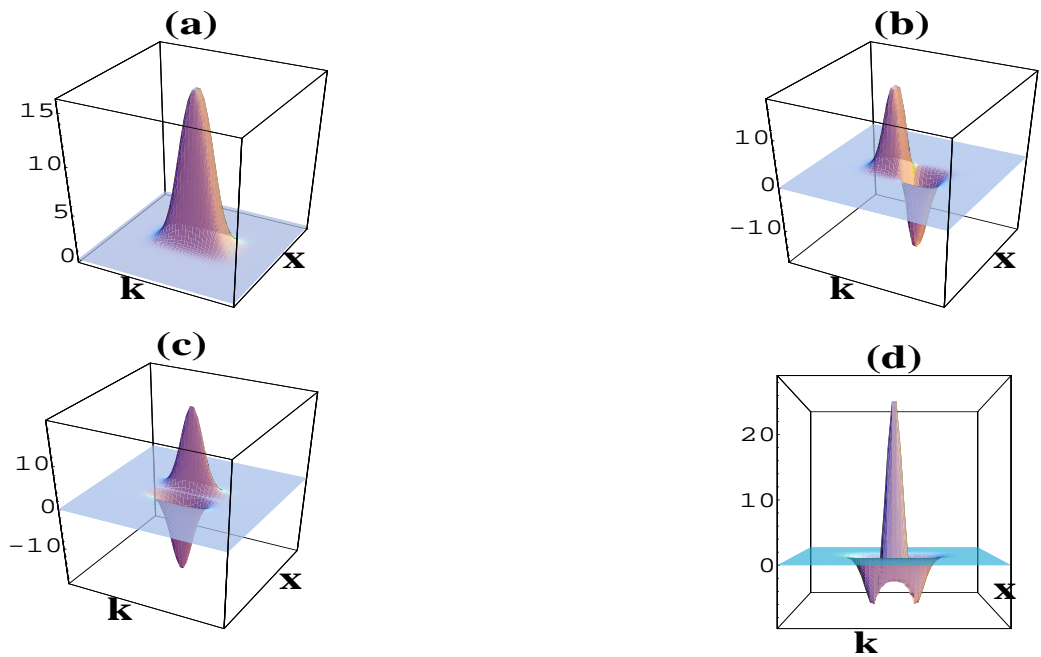,width=16.0cm,height=16.0cm}
\end{center}
\caption{The wavefunctions corresponding to the lowest four
eigenvalues of the reduced model as a function of $x$ and $k$. 
The parameters are 
$g=.2$, $\eta=.00001$,
$m=1$, $\kappa=10$, $n_1=46$, $n_2=74$. 
 (a) Lowest state, (b) first excited state, (c) second
excited state, (d) third excited state. The first and second excited states
should be degenerate in the absence of violation of rotational symmetry.
\label{fig2}}
\end{figure}
\end{document}